\documentclass[a4paper]{jpconf}
\usepackage{graphicx}
\usepackage{wrapfig}
\begin{document}
\title{Micromegas for imaging hadronic calorimetry
}

\author{C.~Adloff, J.~Blaha, S.~Cap, M.~Chefdeville, A.~Dalmaz, C.~Drancourt,
A.~Espargili\`{e}re, R.~Gaglione, R.~Gallet, N.~Geffroy, J.~Jacquemier, 
Y.~Karyotakis, F.~Peltier, J.~Prast and G.~Vouters}
 
\address{Laboratoire d'Annecy-le-Vieux de Physique des Particules, 
Universit\'{e} de Savoie, CNRS/IN2P3, 9 Chemin de Bellevue 74980 
Annecy-le-Vieux, France}

\ead{jan.blaha@lapp.in2p3.fr}

\begin{abstract}
The recent progress in R\&D of the Micromegas detectors for hadronic calorimetry 
including new engineering-technical solutions, electronics development, and 
accompanying simulation studies with emphasis on the comparison of the physics 
performance of the analog and digital readout is described. The developed prototypes 
are with 2 bit digital readout to exploit the Micromegas proportional mode and thus 
improve the calorimeter linearity. In addition, measurements of detection efficiency, 
hit multiplicity, and energy shower profiles obtained during the exposure of small 
size prototypes to radioactive source quanta, cosmic particles and accelerator beams 
are reported. Eventually, the status of a large scale chamber ($1\times1$~m$^2$) are 
also presented with prospective towards the construction of a 1~m$^3$ digital 
calorimeter consisting of 40 such chambers.
\end{abstract}

\section{Introduction}
Calorimetry for experiments at a future $e^+e^-$ collider will use Particle 
Flow~\cite{pfa} as a base technique for the physics event reconstruction. By this 
approach a jet energy resolution lower than 3.8\% for 40 -- 400~GeV jets can be reached 
and thus the accurate separation of $W$ and $Z$ hadronic decays will be possible. For 
optimal performance of this technique, calorimeters with very fine lateral and 
longitudinal segmentation must be employed. One of the proposed technology is a 
Micromegas (micro-mesh gaseous structure) detector with 1~cm$^2$ pads that are readout 
by embedded digital electronics processing only 1 or 2 bit information per channel. 
This concept, which is also called digital calorimetry, makes easier the construction 
of a high granular sampling hadron calorimeter with single shower imaging capability. 
Feasibility of the concept will be validated by a performance test of a 1~m$^3$ 
calorimeter prototype consisting of 40 layers, each equipped with 1~m$^2$ large 
Micromegas chamber, and having about 370,000 readout channels in total. Main outputs 
of the R\&D program towards the construction of this 1~m$^3$ calorimeter prototype are 
described in what follows.

\section{Micromegas basic performance}
Several Micromegas chambers have been produced so far by use of the bulk Micromegas 
technique~\cite{bulk}. Their size varies from $8\times8$~cm$^2$ up to 
$48\times32$~cm$^2$. Each chamber consists of a PCB with top conductive layer segmented
into 1 cm$^2$ anode pads, woven mesh supported by insulating pillars 128~$\mu$m above 
the anode plane, 3~mm thick plastic frame which defines the drift region, and 2~mm 
thick grounded steel cover with a drift copper electrode glued on its internal side. 
Gas distribution is provided by two holes for input and output in the plastic frame. A 
detailed description of the prototype design can be found elsewhere~\cite{micromegas}.

The first prototypes have been equipped with an analog readout which was intended 
mainly for chamber characterization and definition of necessary parameters for future 
prototypes with digital readout chips embedded on the external side of the chamber 
PCB. Three 64-channel ASICs providing digital readout were designed:
HARDROC~\cite{hardroc}, DIRAC~\cite{dirac}, and MICROROC~\cite{microroc}. They can 
operate in a power pulsing mode which is intended for matching the ILC beam time 
structure and therefore to reduce the power consumption down to 10\,$\mu$W/channel. 
The first two chips have been already used for prototype readout and their performance 
has been evaluated during test beam campaigns. A design of the last chip, which will 
equip future chambers, has been completed and its production has just started.

\subsection{Prototype characterization using $^{55}$Fe X-rays}
Fundamental characteristics of the prototype chambers, such as energy resolution,  
electron collection efficiency, gas gain and signal dependency on the environmental 
conditions, were measured at laboratory by use of an $^{55}$Fe X-ray source. The 
measurements were performed in Ar/iC$_4$H$_{10}$ (95/5) and Ar/CO$_2$ (80/20) gas 
mixtures and the signals from X-rays were processed by the analog readout.

The energy resolution of the chamber $\sigma_E/E$ was measured to be 7.5~\% at 5.9~keV
which corresponds to a FWHM of 17.6~\%. The maximum electron collection efficiency is 
reached for an optimal ratio between amplification and drift electric field. It was 
measured that the maximum of the signal is obtained for the field ratio about 150--200 
for both gas mixtures. At this field ratio, the collection efficiency should be maximum.
A maximum gas gain of $4\times10^4$ and $10^4$ has been achieved in Ar/iC$_4$H$_{10}$ 
(95/5) and Ar/CO$_2$ (80/20) gas mixtures, respectively. In addition, the gas gain 
depends strongly on the amplification gap thickness. It is estimated from test beam 
data analysis that gap size irregularities are smaller than 3~$\mu$m r.m.s.. Finally, 
the sensitivity of the gas gain to the ambient variables were measured and quantified 
as $(-1.37\pm0.01)$\%/K$^{-1}$ and $(-0.61\pm0.01)$\%/mbar$^{-1}$ for temperature and 
atmospheric pressure, respectively. These values correspond well to ones predicted 
from a gas gain model~\cite{env}.

\subsection{Prototype test in particle beams}
During the last years the chambers prototypes were tested in particle beams. 
The major objectives of these experiments were the functional test of new prototypes 
in real data taking condition and evaluation of the main chamber properties such as
gain uniformity, detection efficiency and hit multiplicity, and study of the 
prototype response in hadronic and electromagnetic showers. The test beams were 
carried out at CERN PS and SPS lines where chambers were exposed to low momentum 
electrons of 2~GeV/c and hadrons up to 10~GeV/c as well as to high momentum muons and 
pions of 200~GeV/c. All measurements were performed in an Ar/iC$_4$H$_{10}$ (95/5) gas
mixture.

\begin{figure}[htb]
\begin{minipage}{18pc}
\includegraphics[width=14pc, height=13pc]{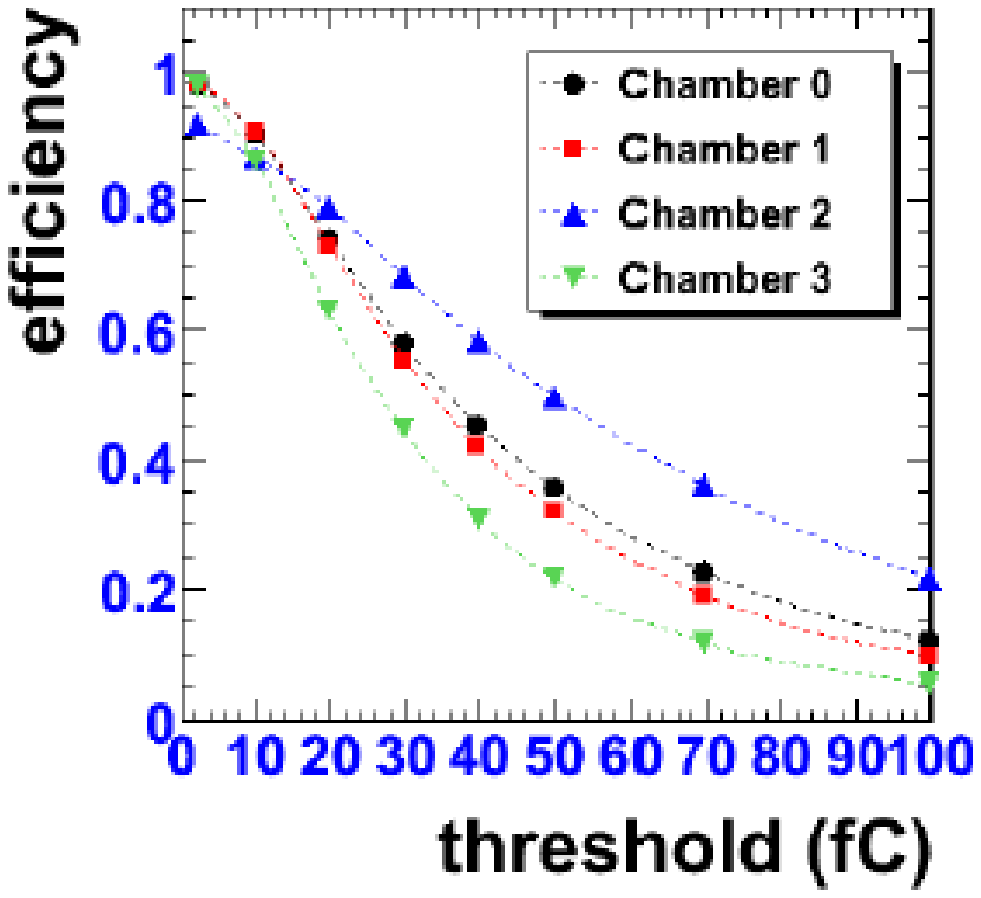}
\caption{\label{efficiency}Detection efficiency as a function of the electronics 
readout threshold.}
\end{minipage}\hfill
\begin{minipage}{18pc}
\includegraphics[width=14pc, height=13pc]{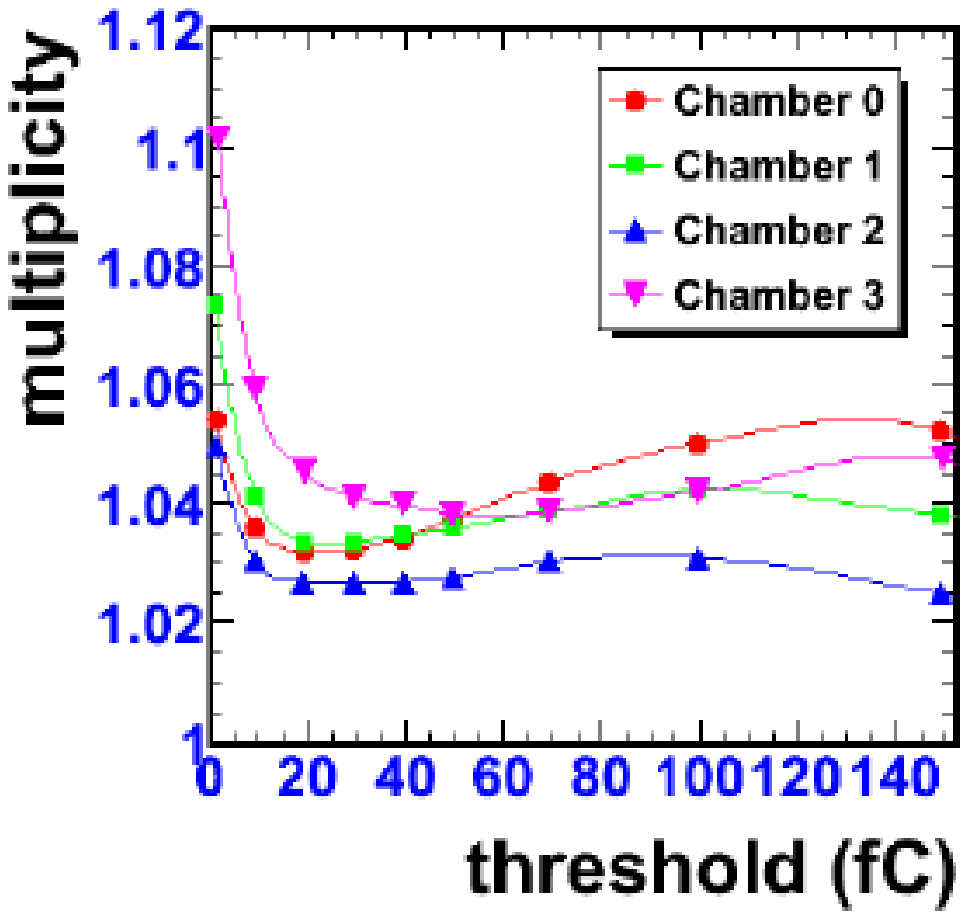}
\caption{\label{multiplicity}Hit multiplicity as a function of the electronics 
readout threshold.}
\end{minipage} 
\end{figure}

\subsubsection{Prototype with analog readout}
Three chambers with an active area of $16\times6$~cm$^2$ and one with 
$32\times12$~cm$^2$, with analog readout were assembled in a stack placed 
perpendicularly to the beam. An electronics noise of 0.3~fC corresponding to 
2000~$e^-$ was measured. The most probable value of the deposited charge is around 
22~fC with a variation of 11.3~\% over all 672 readout channels. Relatively large 
dispersion in gain among the channels is mainly due to drift or/and amplification 
gap non-uniformity. A detection efficiency larger than 97~\% was reached at 1.5~fC 
readout threshold in three of four chambers with an efficiency disparity of less 
than 1~\% per channel. The last one had efficiency around 91~\% as a consequence 
of a higher electronic noise. The efficiency is decreasing with increasing readout 
threshold as is shown in Fig.~\ref{efficiency}. The hit multiplicity was measured to 
be between 1.06 to 1.1 at 1.5~fC readout threshold. Its dependency on the threshold 
is displayed in Fig.~\ref{multiplicity}, where the raised values of multiplicity above 
50~fC are due to the hits from long range $\delta$-rays. 

\begin{figure}[htb]
\centering
\includegraphics[width=17pc]{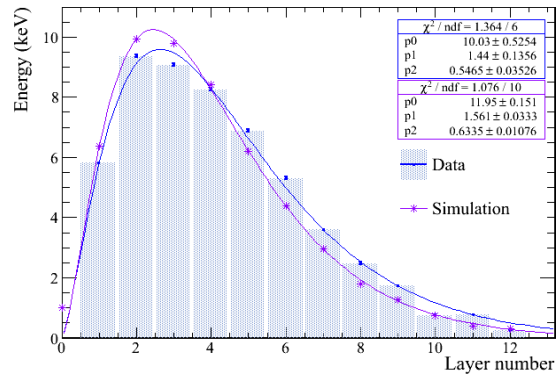}
\caption{\label{profile}Shower profile for 2~GeV/c electrons measured by one
Micromegas chamber and comparison with the Geant4 simulation.}
\end{figure}

In order to verify that the prototypes also detect multi-hits events, shower
profiles have been measured with the $32\times12$~cm$^2$ chamber and a variable 
number of 2~cm thick stainless steel absorber plates placed in front of it. This 
configuration allows to measure lateral and longitudinal shower profiles and also, 
by applying a threshold on the recorded charge signals, to compare the deposited 
energy and number of hits in a chamber. An example of the measured longitudinal 
shower profile for 2~GeV/c electrons and its comparison with Geant4 Monte Carlo 
data is shown in Fig.~\ref{profile} and more details on this study are available 
in Ref.~\cite{prof}.

\subsubsection{Prototype with embedded digital readout}
Two stacks with embedded digital readout were assembled and exposed to 2~GeV/c 
electrons. The first one consists of four chambers with $32\times8$~cm$^2$ pads read
out by HARDROC chips. A clear image of the beam profile has been seen in each chamber. 
Because of the short shaping time of the HARDROC chip, only little of the signal can 
be seen and, consequently, the measured efficiency is about 6~\%. The second stack was 
assembled by four $8\times8$~cm$^2$ chambers with DIRAC chips. Unfortunately, due to 
insufficient spark protection, the life-time of the DIRAC chips in a beam was limited. 
Nevertheless, an efficiency up to 50\% and a hit multiplicity between 1.06 and 1.13 
have been measured. These results are affected by low statistics and therefore serve 
only as a first performance indication. That is why one of the main aim of the next 
generation readout electronics development is to obtain a reliable spark protection. 

\section{1\,m$^2$ prototype: design and sub-detector test}
\begin{figure}[htb]
\begin{minipage}{18pc}\vspace{1.5pc}
\includegraphics[width=17pc]{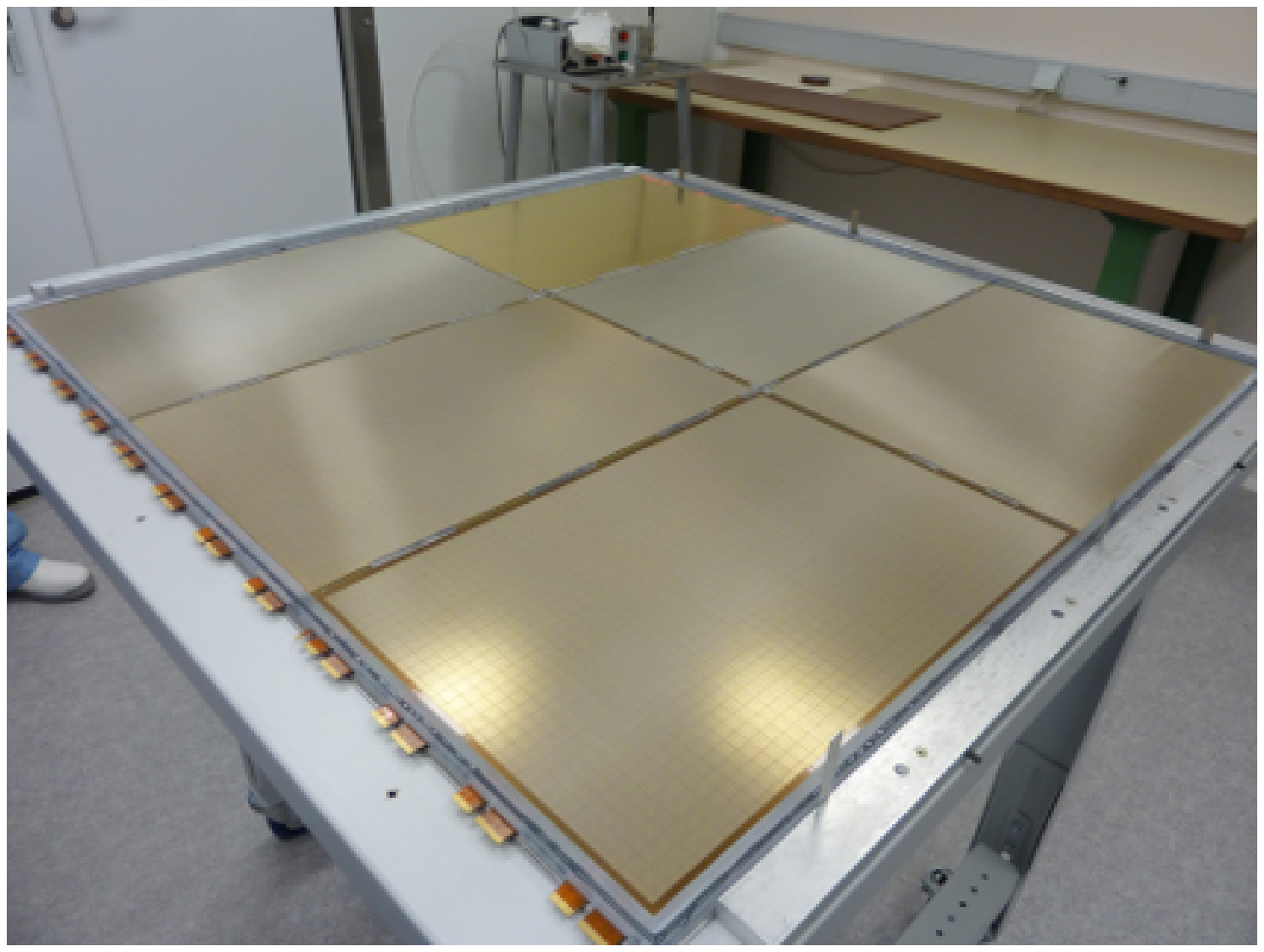}
\caption{\label{m2Photo}The 1\,m$^2$ Micromegas chamber without top steel cover 
showing mesh side of the ASUs.}
\end{minipage}\hfill
\begin{minipage}{18pc}
\includegraphics[width=15pc, height=14pc]{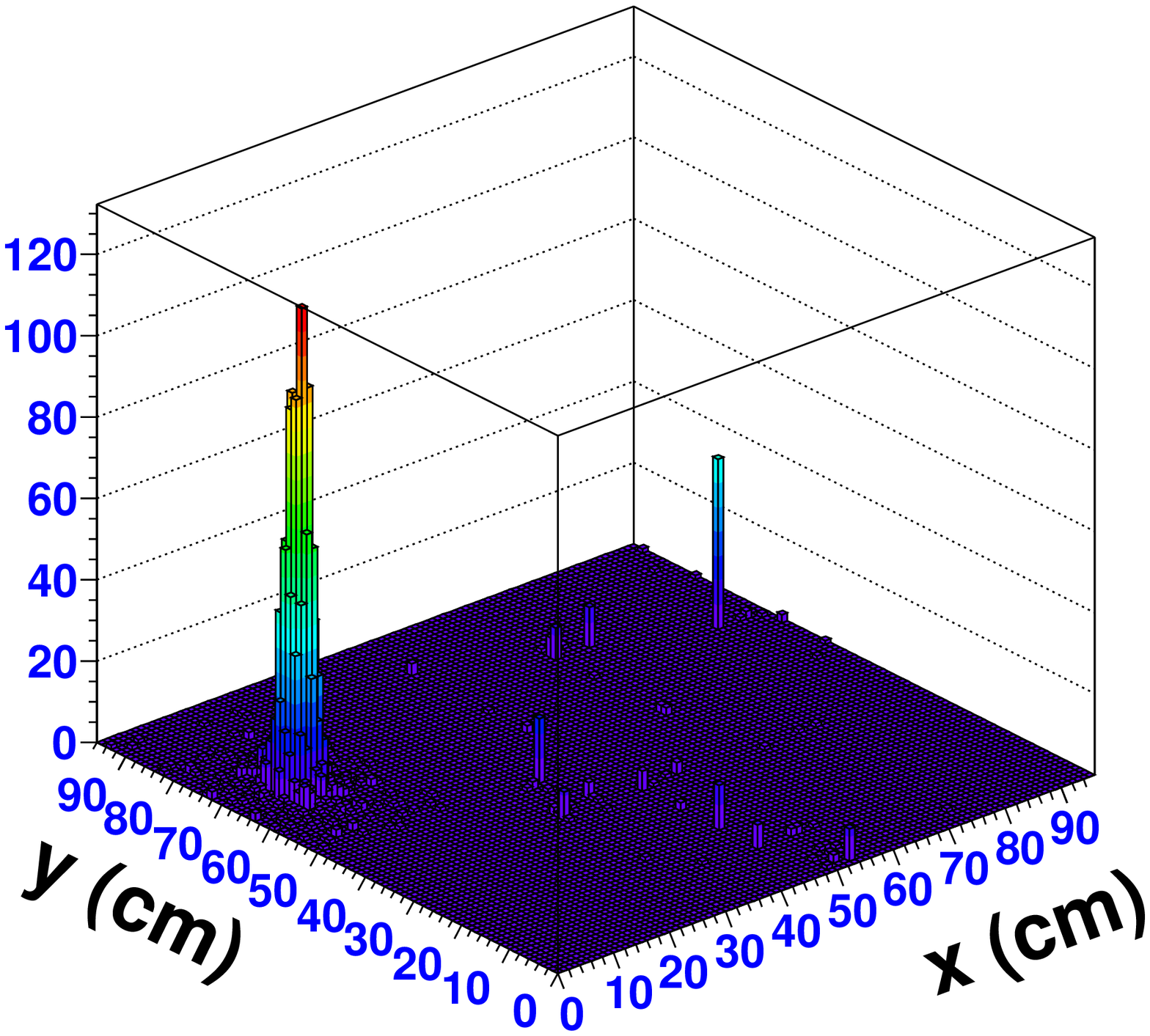}
\caption{\label{beamProfile}A beam profile of 150~GeV/c muons traversing the 1\,m$^2$ 
Micromegas chamber.} 
\end{minipage} 
\end{figure}

An essential step towards the 1\,m$^3$ technological prototype of DHCAL, which will be 
made of 40 Micromegas chambers with embedded readout electronics interleaved by 
stainless steel absorbers, is the construction and test of a 1\,m$^2$ prototype, the 
largest Micromegas chamber so far built. The chamber is formed by 6 active sensor 
units (ASU) of 48$\times$32 readout pads of 1\,cm$^2$, see Fig.~\ref{m2Photo}. Each 
ASU is equipped with a mesh and 24 HARDROC readout chips. These ASUs are glued on a 
2~mm thick stainless steel supporting plate and the chamber is closed by another 2~mm 
thick stainless steel plate that carries the cathode. In this arrangement, all the 
ASUs share the same gas volume. The drift gap is defined by a plastic frame around 
the chamber and spacers that are placed between the ASUs resulting in very small dead 
zones of around 1.6\,\%. The gas distribution is provided by one inlet and outlet in 
the chamber frame. The readout of two ASUs is chained serially and connected to the 
data acquisition system by a detector interface board (DIF). The DIF is a mezzanine 
board which allows to configure HARDROCs by setting readout thresholds and preamplifier 
gains, to readout the HARDORC data, and also to provide the distribution of the system 
clock and high voltages through another dedicated board, called inter-DIF, which stands 
between the DIF and the ASU. The total active area of the complete 1\,m$^2$ chamber 
is 96$\times$96\,cm$^2$ corresponding to 9216 readout channels. The channel signals are 
processed by 144 HARDROC chips and three DIF boards. The effective thickness of the 
chamber, if one does not count the 4 mm of the supporting plates material that can be 
considered as a part of the calorimeter absorber, is equal to 8\,mm and complies well 
with the ILC specification. 

\subsection{ASU test and electronics calibration}
Prior to the 1\,m$^2$ assembly, the electronics characteristics of the readout chips 
are measured and the overall ASU functioning is verified in a test gas chamber.
The electronics tests consist of the determination of the pedestals and noise level 
of each channel to achieve the lowest detection threshold, and the determination of the 
channel preamplifier gains.

The determination of the pedestals and noise level is done by measuring the trigger 
efficiency of each channel as a function of threshold (so-called S-curve). As the 
threshold is common to all chip channels, it is determined by the S-curve with the 
highest inflexion point. By changing the preamplifier gain of each chip channels, it 
is possible to squeeze the S-curves close together and lower the detection threshold 
to minimum (typically down to 3--4~fC).

\begin{figure}[h]
\begin{minipage}{18pc}
\includegraphics[width=15pc, height=14pc]{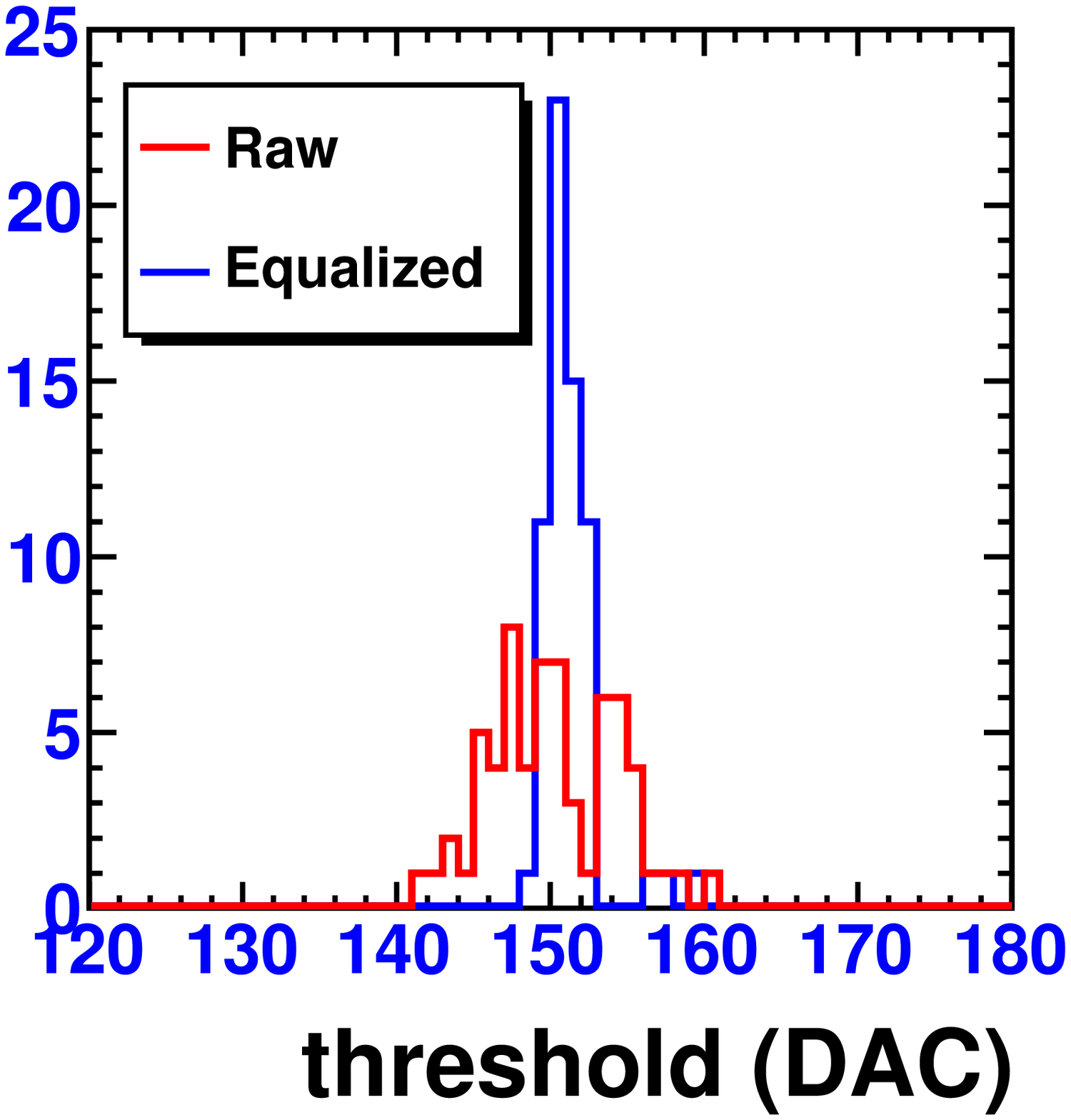}
\caption{\label{calibration}Distributions of the S-curve inflexion points for one 
HARDROC channel before and after the gain equalization for 100~fC input charge.}
\end{minipage}\hfill
\begin{minipage}{18pc}
\includegraphics[width=15pc, height=14pc]{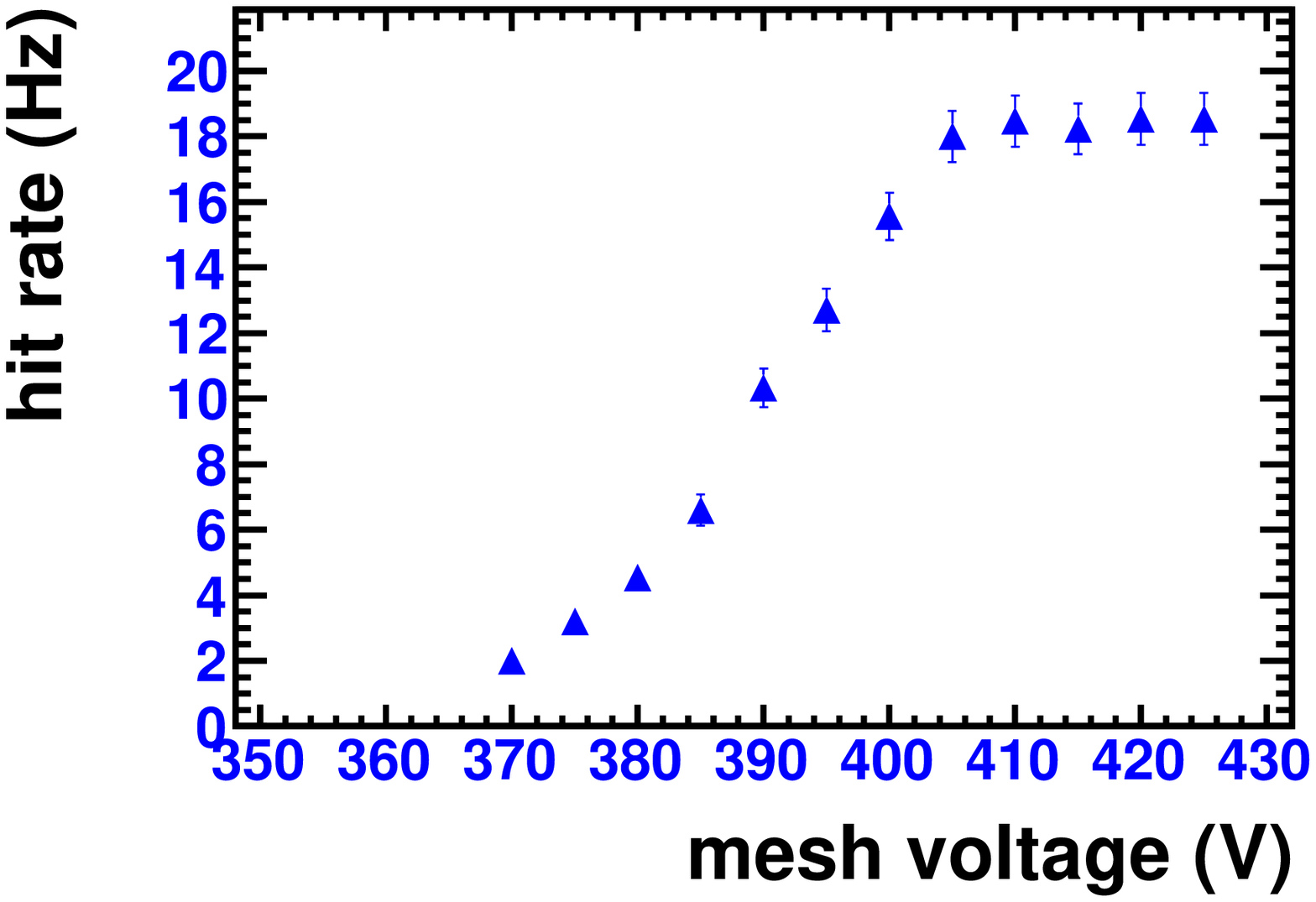}
\caption{\label{hitsVsVmesh}Hit rate from $^{55}$Fe quanta as a function of the voltage
on the mesh. The response is measured for one channel in a gas test box having 3\,cm 
drift gap.}
\end{minipage} 
\end{figure}

The determination of the channel preamplifier gain is performed by injecting 
different test charges in the HARDROC circuitry. Once the gain distribution per 
chip is known, equalization constants are calculated to reduce the spread. The 
equalization procedure has been validated as shown in Fig.~\ref{calibration} where the 
S-curve inflexion point distributions for 100 fC test charge with and without equalized 
gains are plotted.

A dedicated gas chamber, which can house one ASU, has been fabricated to perform 
tests of the ASUs with $^{55}$Fe X-rays before they are mounted into the 1\,m$^2$
chamber. It has a perforated aluminum cover onto which is glued a cathode foil with a 
few nm thin conductive deposit on one side. It is therefore transparent to $^{55}$Fe 
quanta which are almost all absorbed in the 3~cm thick drift gap. By collimating 
the X-ray beam and by measuring the number of hits recorded per unit time on one pad, 
relative variations of efficiency can be inferred. The impact of the chip settings 
and detector voltages can be quantified this way. As an example, the strong influence 
of the mesh voltage (gas gain) on the detection rate is depicted in 
Fig.~\ref{hitsVsVmesh}. Moreover, the box allows to perform the high voltage burn-in 
in air, which effectively reduces ASU's sparking rate by washing out the impurities 
remaining in the amplification gap after the bulk production.

\subsection{The of the first 1\,m$^2$ in a beam}
The first 1\,m$^2$ prototype has been successfully built and tested in a particle beam 
in summer this year. 
This experiment has provided precious information for future mass production. During 
one month operation in a beam, the functionality of the DAQ system and chamber readout 
electronics has been verified. Gas gains up to 15.000 has been reached with very few 
high voltage trips and the first test in a power pulsing mode of the readout 
electronics has been done successfully. The data taken during this period are currently
under analysis. In Fig.~\ref{beamProfile} is shown a beam profile of 150~GeV/c muons 
traversing 1\,m$^2$ Micromegas prototype which is the first sign of the project 
feasibility and an important milestone in the large area Micromegas R\&D effort.  


\section{Simulation studies}
\begin{figure}[htb]
\begin{minipage}{18pc}
\includegraphics[width=17pc]{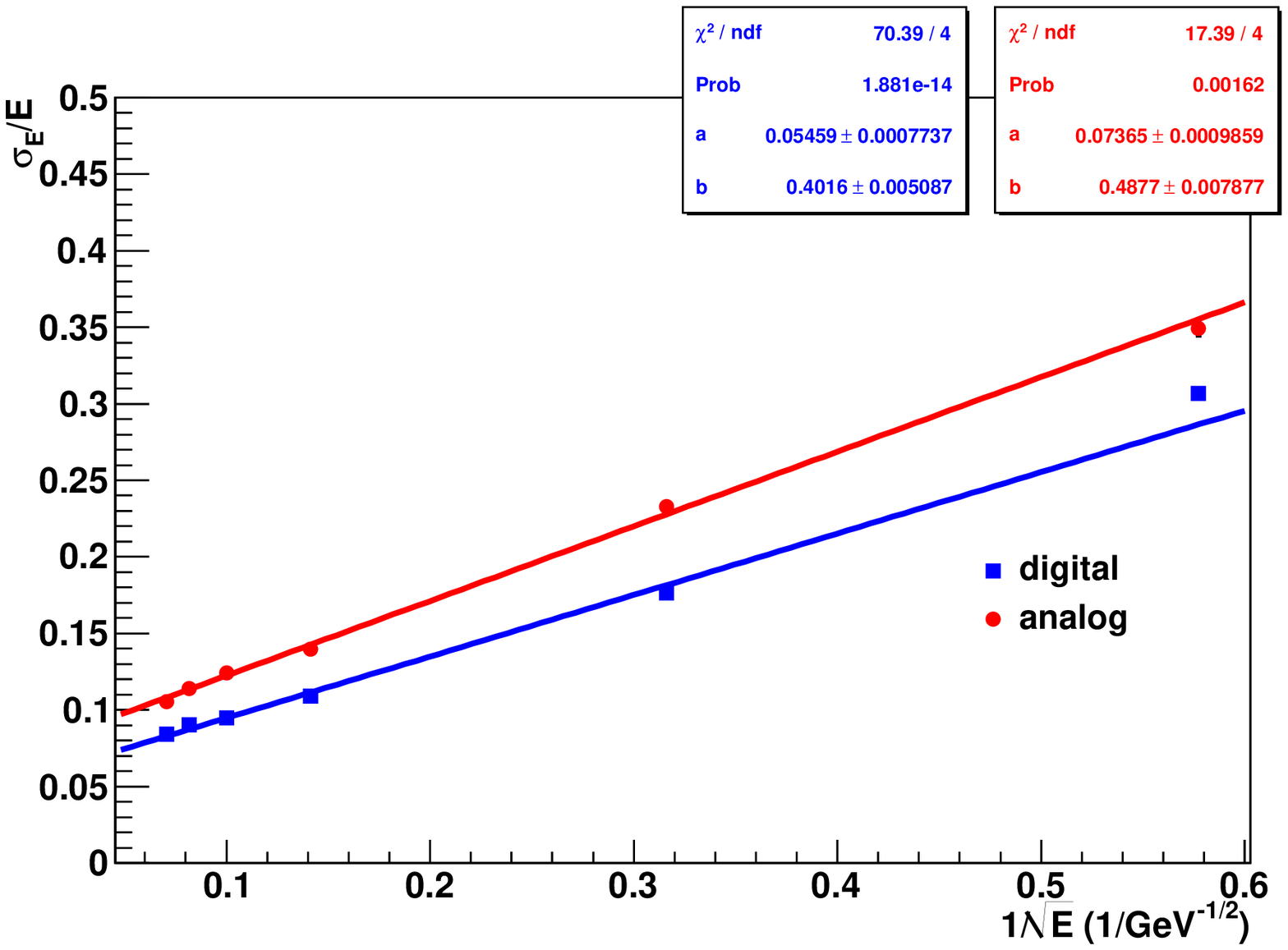}
\caption{\label{engRes}Energy resolution as a function of pion energy for digital 
  and analog readout.}
\end{minipage}\hfill
\begin{minipage}{18pc}
\includegraphics[width=17pc]{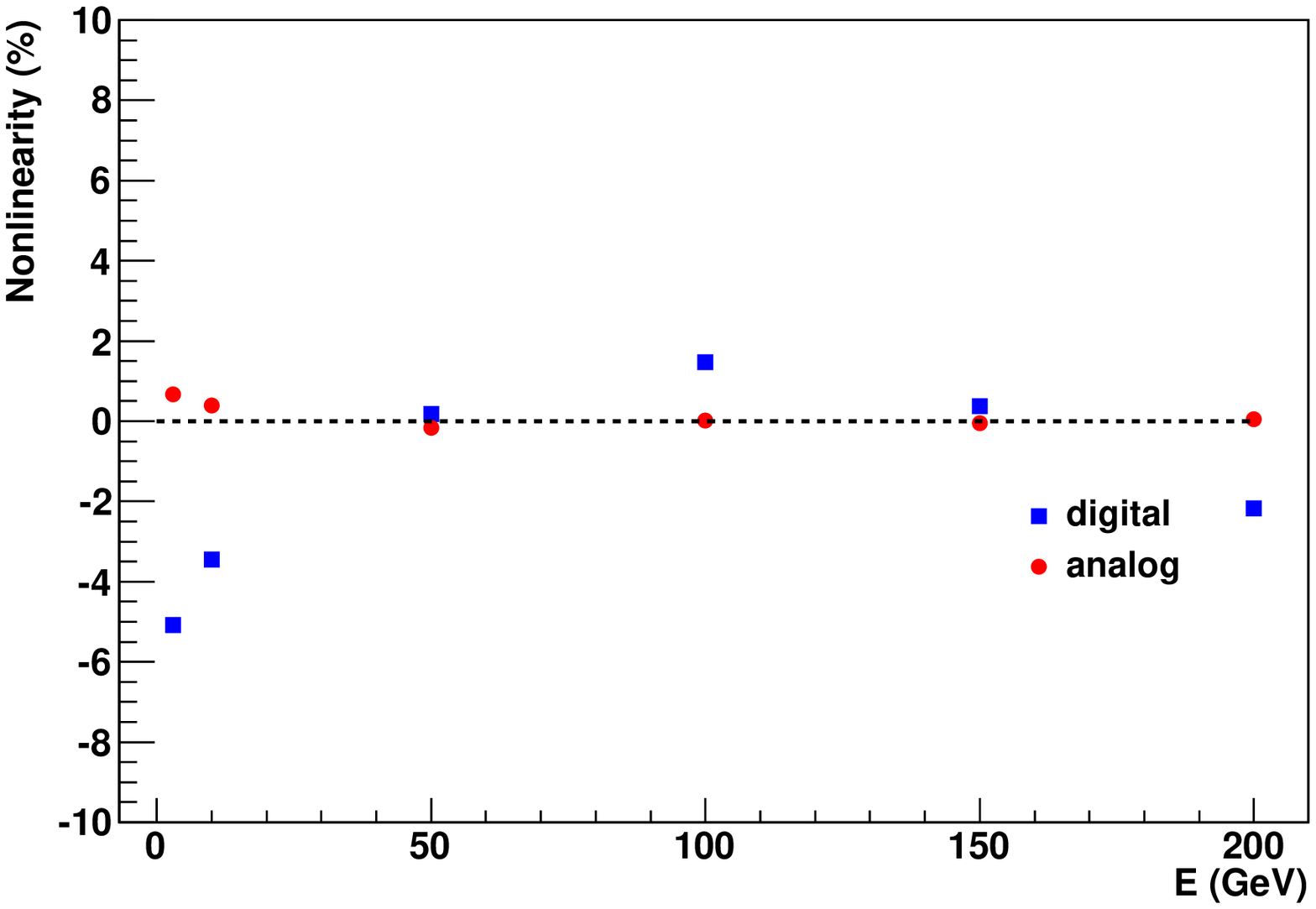}
\caption{\label{linearity}Non-linearity as a function of the pion energy
  for digital and analog readout}
\end{minipage} 
\end{figure}

In order to have better understanding of the physics performance of a digital hadron 
calorimeter with very fine segmentation and its capability to separate hadronic showers,
detailed Monte Carlo simulations of a realistic SiD-like calorimeter module~\cite{sid} 
have been performed~\cite{simulations}. These studies have been focused on the 
calorimeter response and linearity, energy resolution, shower profiles, and the impact 
of the leakage. Results obtained for digital readout with different thresholds were 
compared against the standard analog readout. Fig.\ref{engRes} depicts the energy 
resolution for single pions as a function of their energy for analog and digital 
readout with a readout threshold of 15~fC. Superior energy resolution observed for 
digital readout is due to the suppression of the Landau and particle path length 
fluctuations. On the other hand, due to the saturation in calorimeter response, the 
linearity of a digital calorimeter is not as good as it is in case of analog one, see 
Fig.\ref{linearity}. The saturation effect can be reduced by use of semi-digital 
readout with two or three thresholds. The determination of the optimal thresholds and 
their corresponding weights is the subject of current investigation.

\section{Conclusions}

Several Micromegas chambers have been developed for an application in digital hadron 
calorimetry. The prototypes with analog readout have been characterized and extensively 
tested with different particle sources. The tests of these chambers have shown very 
good performance on detection efficiency and hit multiplicity, the parameters which 
are crucial for calorimetry at future linear colliders. Based on the obtained results, 
new generation Micromegas chambers with embedded readout electronics have 
been designed and produced. The feasibility of the project has been proven by 
construction and functionality test of the 1\,m$^2$ prototype in particle beams. The 
simulation studies of the digital hadronic calorimeter are well advanced.   


\section{References}

\end{document}